\documentclass[a4paper]{article}
\usepackage[english]{babel}
\input{epsf}
\usepackage{mathtext}
\usepackage{graphics,dcolumn,bm,float}
\usepackage{amssymb,amsmath,rotate,color}
\usepackage{times}
\usepackage{color}
\usepackage{fancyhdr}
\newcommand{\be}{\begin{equation}}
\newcommand{\ene}{\end{equation}}
\newcommand{\ba}{\begin{array}}
\newcommand{\ea}{\end{array}}

\begin{document}
\title{Super quantum Dirac operator on the q-deformed super fuzzy sphere $ S_{q\mu}^{(2\vert2)} $ in instanton sector using quantum super Ginsparg-Wilson algebra}
\author{M. Lotfizadeh\thanks{E-mail: M. Lotfizadeh@urmia.ac.ir}$^{\;\;,1}$  \\
\footnotesize\textit{$^1$Department of Physics, Faculty of Science, Urmia University, P.O.Box: 165, Urmia, Iran}\\
\footnotesize\textit{}}
\date{}
\maketitle
\begin{abstract}
It has been constructed the quantum super fuzzy Dirac and chirality operators on q-deformed super fuzzy  sphere $ S^{(2\vert2)}_{q\mu} $. Using the quantum super fuzzy Ginsparg-Wilson algebra, it has been studied the q-deformed super gauged fuzzy Dirac and chirality operators in instanton sector. It has been showed that they have correct commutative limit in the limit case when noncommutative parameter $ l $ tends to infinity and $ q\rightarrow 1 $.
\end{abstract}
\textbf{PACS}: 74.45.+c; 85.75.-d; 73.20.-r\\
\textbf{Keywords}: q-deformed super fuzzy sphere, gauged quantum super Ginsparg-Wilson algebra, q-deformed gauged super Dirac and chirality operators.

\section{Introduction}

Dirac and chirality operators are two important self-adjoint operators for the Connes-Lott approach to noncommutative geometry. A unital spectral triple [1-2], $ (\mathcal{A},\mathcal{H},D) $ consists of a complex unital $ \ast $-algebra $ \mathcal{A} $, faithfully $ \ast $-represented by bounded operators on a separable Hilbert space $\mathcal{H} $, and a self-adjoint operator $ D: \mathcal{H}\rightarrow \mathcal{H} $ (the Dirac operator) with the following properties:

$ \bullet $the resolvent $ (D-\lambda)^{-1},\,\lambda \notin \mathbb{R} $ is a compact operator on $\mathcal{H} $, 

$ \bullet $for all $ a \in \mathcal{A} $ the commutator $ [D, \pi(a)] $ is a bounded operator on $\mathcal{H}$.

A spectral triple $ (\mathcal{A},\mathcal{H},D) $ is called even if there exists a $ \mathbb{Z}_{2} $-grading of $\mathcal{H}$, i.e. An operator $ \gamma :\mathcal{H}\rightarrow \mathcal{H}$ with $ \gamma^{\ast}=\gamma $ and $ \gamma^{2}=1$, such that $ \gamma D+D\gamma=0 $ and $ \gamma a=a \gamma $ for all $ a \in \mathcal{A} $. Otherwise the spectral triple is said to be odd. For odd dimensional manifolds, there are no chirality operators and in such a case, the Dirac operator describes only the differential structures. There are three types of Dirac and chirality operators on the fuzzy two-sphere. Ginsparg-Wilson Dirac operator, $ D_{GW} $ [3-10], Watamura-Watamura Dirac operator $ D_{WW} $ [11-13] and Grosse-Klimcik-Presnajder Dirac operator $ D_{GKP} $ [14-15]. These three types of Dirac operators are compared with each other in [16].
The fuzzy super sphere $ S_{F}^{(2\vert2)} $ was extensively studied in [17,18]. It has been showed that the $ S_{F}^{(2\vert2)} $ is described by the irreducible representations of the Lie superalgebra $ uosp(2\vert1) $. After that, many studies on different aspects of field theory on $ S_{F}^{(2\vert2)} $ have been caried out [19-25]. The star product on the $ S_{F}^{(2\vert2)} $ has been studied in [19]. In[20] non linear sigma model on $ S_{F}^{(2\vert2)} $ has been studied. Gauge theory on $ S_{F}^{(2\vert2)} $ from a supermatrix model has been investigated in [21,22]. Graded Hopf fibration and its corresponding super monopole has been studied in [23,24,25].
The idea of q-deformed geometry was extensively studied in the late 1980 s and 1990 s. The q-deformed Hopf fibration has been studied in the frame work of Hopf-Galois extension in [27]. Podles sphere are introduced in [28-29]. The q-deformed Dirac operator on quantum Podles sphere has been studied from different approaches [30-35]. The q-deformed Watamura-Watamura Dirac operator $ D_{WW}^{q} $, has been studied in [30]. The authors were constructed Dirac and chirality operators on noncommutative space having $ U_{q}(su(2))$ as the symmetry group. It has been argued that the Dirac operator is covariant and in the commutative limit where the underlying space is Podles sphere, the full rotational invariance of the Dirac operator is recovered. It was further shown that the Dirac operator reduces to that obtained in [12]. In [32] the goal of the authors is to describe q-deformed version of GKP Dirac operator on quantum sphere.In [5] it has been showed that how one can construct gauged Dirac operator in instanton sector on fuzzy sphere. To construct the q-deformed Dirac operator on quantum fuzzy super sphere we need the quantum deformation of the superalgebra $ osp_{q}(1\vert2) $. In [36,37] a quantum analogue of the simplest super algebra  $ osp_{q}(1\vert2) $ and its finite dimensional irreducible representations have been studied. In [38], quantum super 2-spheres and the corresponding quantum super transformation group are introduced.
 In this paper we generalize Ginsparg-Wilson algebra to quantum super fuzzy Ginsparg-Wilson algebra and construct its super Dirac and chirality operators. This paper is organized as follows: In section 2 we briefly study the super sphere, fuzzy super sphere and q-deformed fuzzy super sphere. In section 3, it has been studied the quantum graded Hopf fibration and its projective module construction. In section 4, spin $\frac{1}{2}$ q-deformed superprojectors of the left and right superprojective $ A(S_{q\mu}^{(2\vert2)}) -$module have been constructed.
Quantum super fuzzy Ginsparg-Wilson algebra and its superSpin $\frac{1}{2}$ fuzzy q-deformed super Dirac and chirality operators have been studied in section 5. In section 6, quantum super gauged Dirac and chirality operators was constructed. In section 7, instanton coupling and in section 8, gauging the quantum super fuzzy Dirac operator in instanton sector have been studied, respectively.

\section{Supersphere, fuzzy supersphere and q-deformed fuzzy super sphere}

The superspace $ R^{(3\vert2)} $ is defined as the algebra of polynomials in generators $ x_{i} $ and $ \theta_{\alpha} $ satisfying reality conditions
\begin{equation}
x_{i}^{\ddagger}= x_{i}, \quad \theta_{\alpha}^{\ddagger}= \varepsilon_{\alpha\beta}\theta_{\beta}.\tag{2-1}
\end{equation}
The equation characterizing the adjoint orbit $ S^{(2\vert2)} $ of $ UOSp(2\vert1) $ is
\begin{equation}
S^{(2\vert2)}=\langle (x_{i}, \theta_{\alpha})\in R^{(3\vert2)} \quad| \quad x_{i}x_{i}+ \varepsilon_{\alpha\beta}\theta_{\alpha}\theta_{\beta}=1\rangle\tag{2-2}
\end{equation}
The $ uosp(2\vert1)$ algebra contains the $ su(2) $algebra as its maximal bosonic subalgebra and consists of five generators there of which are bosonic $ L_{i}(i=1,2,3) $and two of which are fermionic $ L_{\alpha}(\alpha= +,-) $. They satisfy:
\begin{equation}
[L_{i}, L_{j}]= i\varepsilon_{ijk}L_{k},\quad [L_{i}, L_{\alpha}]= 1/2(\sigma_{i})_{\beta\alpha}L_{\beta},\quad \lbrace L_{\alpha}, L_{\beta}\rbrace = 1/2 (\varepsilon\sigma_{i})_{\alpha\beta}L_{i}\tag{2-3}
\end{equation}
where $ \varepsilon= i\sigma_{2} $. The even part of this algebra is $ su(2) $, which is generated by $ L_{i} $, and the odd generators $ L_{\alpha} $ are $ su(2) $ spinors. The irreducible representations of $ uoso(2\vert1) $ algebra are characterized by the Casimir operator
\begin{equation}
C_{uosp(2\vert1)}= L_{i}L_{i}+ \epsilon_{\alpha\beta}L_{\alpha}L_{\beta},\tag{2-4}
\end{equation}
and its eigenvalues are given by $ l(l+1/2) $ with $ l $ superspin that takes non-negative integers and half-integers, $ l= 0, 1/2, 1,... $. $ L_{i} $ and $ L_{\alpha} $ are super Hermitian
\begin{equation}
L_{i}^{\ddagger}= L_{i}^{\dagger}= L_{i},\quad L_{\alpha}^{\ddagger}= L_{\alpha},\tag{2-5}
\end{equation}
where $ \ddagger $ is the super-adjoint.
Noncommutative geometry is a pointless geometry. In this geometry instead of the coordinates $ (x_{i}, \theta_{\alpha}) $ of $ S^{(2\vert2)} $, the $ uosp(2\vert1) $ angular momentum generators in the unitary  irreducible $ l-$representation space have the role of the points of the fuzzy $ S^{(2\vert2)} $ i.e $ S_{\mu}^{(2\vert2)} $. 
Let us consider $ X_{i}= \mu L_{i} $ and $ \Theta_{\alpha}= \mu L_{\alpha} $ with $ \mu $ determined by the value of $ uosp(2\vert1) $ Casimir operator 
\begin{equation}
\dfrac{1}{\mu^{2}•}= C_{uosp(2\vert1)}= l(l+1/2)\tag{2-6}
\end{equation}
Then, the noncommutative coordinates 
\begin{equation}
X_{i}= \dfrac{L_{i}}{\sqrt{l(l+ 1/2)}•}, \quad \Theta_{\alpha}= \dfrac{L_{\alpha}}{•\sqrt{l(l+ 1/2)}}\tag{2-7}
\end{equation}
which satisfy the $ uosp(2\vert1) $algebra
\begin{equation*}
[X_{i}, X_{j}]= \dfrac{i}{\sqrt{l(l+1/2)}•}\varepsilon_{ijk}X_{k},
\end{equation*}
\begin{equation*}
 [X_{i}, \Theta_{\alpha}]= \dfrac{1}{2\sqrt{l(l+1/2)}•}(\sigma_{i})_{\beta\alpha}\Theta_{\beta},
 \end{equation*}
 \begin{equation}
 \lbrace \Theta_{\alpha}, \Theta_{\beta}\rbrace = \dfrac{1}{2\sqrt{l(l+1/2)}•} (\varepsilon\sigma_{i})_{\alpha\beta}L_{i}\tag{2-8}
\end{equation}
are determined fuzzy supersphere $ S_{\mu}^{(2\vert2)} $ by relation
\begin{equation}
X_{i}X_{i}+ \varepsilon_{\alpha\beta}\Theta_{\alpha}\Theta_{\beta}=1.\tag{2-9}
\end{equation}
The coordinates $ (x_{i}, \theta_{\alpha}) $ of commutative supersphere can be obtained as the limit case
\begin{equation}
 x_{i}=\lim_{l \to \infty}\frac{L_{i}^{L,R}}{\sqrt{l(l+1/2)}}  =\lim_{l \to \infty}\dfrac{L_{i}^{L,R}}{l}, \quad  \theta_{\alpha}=\lim_{l \to \infty}\frac{L_{\alpha}^{L,R}}{\sqrt{l(l+1/2)}}  =\lim_{l \to \infty}\dfrac{L_{\alpha}^{L,R}}{l}
\tag{2-10}
\end{equation}
In the Hopf fibration  $S^{(3\vert2)}\;\xrightarrow{U(1)}\;S^{(2\vert2)}$, the module of sections is $G(S^{(2\vert2)})$-module $\Gamma^{\infty}(S^{(2\vert2)},E^{(n)})$ in which $G(S^{(2\vert2)})$ is the commutative algebra of superfunctions on the supersphere $S^{(2\vert2)}$. In the fuzzy case, this algebra is a noncommutative algebra and therefore left and right modules are not isomorphic. In this case to each angular momentum operator $\mathbf{L}= (L_{i}, L_{\alpha})$, We associate two linear operators $\mathbf{L}^{L}$ and $\mathbf{L}^{R}$ with the left and right actions on the fuzzy super Hermitian matrix algebra $\mathcal{A}_{l}= \lbrace\psi\in M_{4l+1}(C_{L})\rbrace $:
\begin{equation}
L_{i}^{L} \psi=L_{i} \psi ,\quad L_{i}^{R} \psi=\psi L_{i} ,\quad L_{\alpha}^{L} \psi= L_{\alpha} \psi,\quad L_{\alpha}^{R} \psi= \psi L_{\alpha},\quad\forall\psi \in \mathcal{A}_{l},
\tag{2-11}
\end{equation}
where the right action satisfies the $ uosp(2\vert1) $ algebra with minus sign $ (- L_{i}^{R}, -L_{\alpha}^{R}) $
.These left and right operators commute with each other:
\begin{equation}
[L_{i}^{L},L_{j}^{R}]=0, \quad [L_{\alpha}^{L},L_{\beta}^{R}]=0.
\tag{2-12}
\end{equation}
The $ \mathbf{L}^{L} $ and $ \mathbf{L}^{R} $ have the same $ uosp(2\vert1) $ algebra with the Casimir $ C_{uosp(2\vert1)} $:
\begin{equation}
C_{uosp(2\vert1)}\equiv \mathbf{L}^{L}\cdot  \mathbf{L}^{L}=\mathbf{L}^{R}\cdot \mathbf{L}^{R}=l(l+1/2)1.
\tag{2-13}
\end{equation}
We use $ \mathbf{L}^{L} $, $ \mathbf{L}^{R} $ to define the fuzzy version of orbital momentum operator $\boldsymbol{\mathcal{L}}$ on the fuzzy supersphere $ S_{F}^{(2\vert2)}$. We define $\boldsymbol{\mathcal{L}}$ by the adjoint action of $ L_{i} $ and $ L_{\alpha} $ on the $\mathcal{A}_{l}$:
\begin{equation}
\mathcal{L}_{i} \psi =(L_{i}^{L} - L_{i}^{R})\psi= ad_{L_{i}}\psi = [L_{i},\psi].
\tag{2-14}
\end{equation}
\begin{equation}
\mathcal{L}_{\alpha} \psi =(L_{\alpha}^{L} - L_{\alpha}^{R})\psi= ad_{L_{\alpha}}\psi = [L_{\alpha},\psi].
\tag{2-15}
\end{equation}
Let, $ \mathbf{\sigma_{1}},\mathbf{\sigma_{2}},\mathbf{\sigma_{3}} $, be Pauli matrices of $ su(2) $ Lie algebra. Then, we can define the super version of these matrices as the fundamental superspin $ 1/2 $ representation of $ uosp(2\vert1) $as:
\begin{equation}
	\Sigma_{i}= 1/2
	\begin{pmatrix}
	\sigma_{i} & 0\\
	0 & 0
	\end{pmatrix},\quad
	\Sigma_{\alpha}= 1/2
	\begin{pmatrix}
	0 & \tau_{\alpha}\\
	-(\varepsilon \tau_{\alpha})^{t} & 0
	\end{pmatrix}	
\tag{2-16}
\end{equation}
with $ \varepsilon= i\sigma_{2}, \tau_{1}= (1, 0)^{t} $ and $ \tau_{2}= (0, 1)^{t} $.
Let us define the q- number $ n $ as
\begin{equation}
[n]_{q}= \dfrac{q^{n}-q^{-n}}{q-q^{-1}}.
\tag{2-17}
\end{equation}

Let us denote by $ S_{q\mu}^{(2\vert2)} $ the fuzzy quantum supersphere. Also, we denote the generators of the coordinate algebra of $ S_{q\mu}^{(2\vert2)} $ i.e. $ A( S_{q\mu}^{(2\vert2)} )$ by $ X_{+},X_{-}, X_{0}, \Theta_{+}, \Theta_{-} $, along with the unit 1. Let us define these coordinates as
\begin{equation}
X_{i}= \dfrac{[2]_{q}L_{i}}{\sqrt{[2l]_{q}[2l+ 1]_{q}}•}, \quad \Theta_{\alpha}= \dfrac{[2]_{q}L_{\alpha}}{•\sqrt{[2l]_{q}[2l+ 1]_{q}}}\tag{2-18}
\end{equation}They satisfy the commutation relations:
\begin{equation}
\mathbf{X}_{+}\mathbf{X}_{-}-\mathbf{X}_{-}\mathbf{X}_{+}=\mu \mathbf{X}_{0}-(q-q^{-1})X_{0}^{2},\qquad [\mathbf{X}_{0},\mathbf{X}_{\pm}]_{q}=\pm \mu \mathbf{X}_{\pm},
\tag{2-19}
\end{equation}
with the supersphere constraint
\begin{equation}
\mathbf{X}\cdot \mathbf{X}= \sum_{0, \pm} q^{m} X_{-m}X_{m} + \varepsilon_{\alpha\beta}\theta_{\alpha}\theta_{\beta}= X_{0}^{2}+qX_{-}X_{+}+q^{-1}X_{+}X_{-}+ \varepsilon_{\alpha\beta}\theta_{\alpha}\theta_{\beta}=1.
\tag{2-20}
\end{equation}
The group $ UOSp_{q}(2\vert1) $ is the symmetry group of $ S_{q\mu}^{(2\vert2)} $. The quantum superalgebra $ U_{q}(uosp(2\vert1))= uosp_{q}(2\vert1) $ is the $ Z_{2}- $ graded unital associative algebra generated by the even element $ L_{0} $ and the odd elements $ L_{\pm} $ satisfying the relations 
\begin{equation}
[L_{0}, L_{\pm }]= \pm L_{\pm }, \quad  \lbrace L_{+}, L_{-} \rbrace = [2L_{0}]_{q^{1/2}}.
\tag{2-21}
\end{equation}
The $ uosp_{q}(2\vert1) $ can be concretized by adding the grade involution $ E $ to the set of generators. Thus the quantum superalgebra $ uosp_{q}(2\vert1) $ satisfy the relations
 \begin{equation}
[L_{0}, L_{\pm }]= \pm L_{\pm }, \quad  \lbrace L_{+}, L_{-} \rbrace = [2L_{0}]_{q^{1/2}}, \quad \lbrace E, L_{\pm}\rbrace =0, \quad [E, L_{0}]=0, \quad E^{2}=1.
\tag{2-22}
\end{equation}
 By introducing new generators
 \begin{equation}
 K= q^{L_{0}}, \quad K^{-1}= q^{-L_{0}},\tag{2-23}
 \end{equation}
one can write equations (2-22) as:
\begin{equation*}
 KL_{+}K^{-1}= q L_{+},\quad  KL_{-}K^{-1}= q^{-1} L_{-},\quad KK^{-1}= 1, \quad E^{2}= 1,
 \end{equation*}
\begin{equation}
 [K, E]= 0,\quad [K^{-1}, E]=0,\quad \lbrace L_{\pm}, E \rbrace =0, \quad \lbrace L_{+}, L_{-}\rbrace = \dfrac{K- K^{-1}}{q^{1/2}- q^{- 1/2}•}\tag{2-24}
 \end{equation}
The Casimir operator $ C_{uosp_{q}(2\vert1)} $ which defines as
 \begin{equation}
C_{uosp_{q}(2\vert1)}= (L_{+}L_{-}- [L_{0}- 1/2]_{q})E\tag{2-25}
 \end{equation}
commutes with all the generators in (2-22) and has the eigenvalues $ \dfrac{[2l]_{q}[2l+1]_{q}}{•[2]_{q}^{2}} $ which in the limit case $ q\rightarrow 1 $ reduces to $ l(l+1/2) $. Introducing the coproduct map $ \Delta: uosp_{q}(2\vert1) \rightarrow uosp_{q}(2\vert1)\otimes uosp_{q}(2\vert1) $, counit map $ \epsilon: uosp_{q}(2\vert1)\rightarrow C $ and coinverse map $ \sigma: uosp_{q}(2\vert1)\rightarrow uosp_{q}(2\vert1) $, the Hopf algebra structure of $ uosp_{q}(2\vert1) $ can be given by the relations
\begin{equation*}
 \Delta (L_{+})= L_{+}\otimes KE + 1\otimes L_{+},\quad \Delta (L_{-})= L_{-}\otimes E + K^{-1}\otimes L_{-}
 \end{equation*}
\begin{equation*}
\Delta(K)= K\otimes K, \quad \Delta(E)= E\otimes E
 \end{equation*}
\begin{equation*}
\epsilon(E)= 1,\quad \epsilon(K)=1,\quad \epsilon(L_{\pm })=0,
 \end{equation*}
\begin{equation}
\sigma(E)= E,\quad \sigma(K)= K^{-1},\quad \sigma(L_{+})= -L_{+}K^{-1}E, \quad \sigma(L_{-})= -KL_{-}E.\tag{2-25}
 \end{equation}

In the quantum fuzzy supersphere $ S_{q\mu}^{(2\vert2)} $ we have two different parameter $ q $ and $ \mu $, which we take to be real. Then, there are different limiting cases. These different limits can be expressed as:

$ \bullet $ first limit: $ S_{q\mu}^{(2\vert2)} \longrightarrow  S_{\mu}^{(2\vert2)} \longrightarrow S^{(2\vert2)}   $,\;($ q\longrightarrow 1 $ followed by $ \mu\longrightarrow 0 $)

$ \bullet $ second limit: $ S_{q\mu}^{(2\vert2)} \longrightarrow  S_{q}^{(2\vert2)} \longrightarrow S^{(2\vert2)}   $,\;($\mu \longrightarrow 0 $ followed by $q\longrightarrow 1 $)

$ \bullet $ third limit: $ S_{q\mu}^{(2\vert2)} \longrightarrow  S^{(2\vert2)}   $,\;($\mu \longrightarrow 0 $,\:  $q\longrightarrow 1 $ simultaneously).
\\where $ S_{q\mu}^{(2\vert2)} $ is the fuzzy supersphere. Similar to definition (2-7), using the Casimir (2-25), we define $ \mu_{q} $ as:
\begin{equation}
\mu_{q}\equiv \dfrac{[2]_{q}}{\sqrt{[2l]_{q}[2l+1]}}
\tag{2-26}
\end{equation}
where in the limit $ q\longrightarrow 1 $ it becomes $ \dfrac{1}{\sqrt{l(l+1/2)}} $.

In the third limit case the constraint (2-20) becomes the well known supersphere $ S^{(2\vert2)} $:
\begin{equation}
\mathbf{x}\cdot \mathbf{x}= x_{i}x_{i}+ \varepsilon_{\alpha\beta}\theta_{\alpha}\theta_{\beta}=1.
\tag{2-27}
\end{equation}
Let, $ \mathbf{\sigma_{1}},\mathbf{\sigma_{2}},\mathbf{\sigma_{0}} $, be Pauli matrices of $ su(2) $ Lie algebra. Then, we can define the q-deformed version of these matrices as:
\begin{equation}
\sigma_{1}^{q} = \sqrt{\frac{[2]_{q}}{2q}}\sigma_{1} = \sqrt{\frac{[2]_{q}}{2q}}
	\begin{pmatrix}
	0 & i\\
	i & 0
	\end{pmatrix}, \qquad 
	\sigma_{2}^{q} = \sqrt{\frac{[2]_{q}}{2q}} \sigma_{2} = \sqrt{\frac{[2]_{q}}{2q}} 
	\begin{pmatrix}
	0 & 1\\
	-1 & 0
	\end{pmatrix},\qquad 
	\sigma_{0}^{q} = 
	\begin{pmatrix}
	q & 0\\
	0 & -q^{-1}
	\end{pmatrix}
\tag{2-28}
\end{equation}
Now, we can define the super version of these matrices as
\begin{equation}
	\Sigma_{i}^{q}= \dfrac{1}{[2]_{q}•}
	\begin{pmatrix}
	\sigma_{i}^{q} & 0\\
	0 & 0
	\end{pmatrix},\quad
	\Sigma_{\alpha}^{q}= \dfrac{1}{[2]_{q}•}
	\begin{pmatrix}
	0 & \tau_{\alpha}\\
	-(\varepsilon \tau_{\alpha})^{t} & 0
	\end{pmatrix}	
\tag{2-29}
\end{equation}
with $ \varepsilon= i\sigma_{2}^{q}, \tau_{1}= (1, 0)^{t} $ and $ \tau_{2}= (0, 1)^{t} $.

\section{Module construction}
Consider the $U(1)$ principal fibration  $\pi$ with $S^{(3\vert2)} \cong UOSp(2\vert1)$ as total space:
\begin{equation}
U(1) \;\xrightarrow{right U(1)-action}\; S^{(3\vert2)}\;\xrightarrow{\pi}\; S^{(2\vert2)},
\tag{3-1}
\end{equation}
over the $ (2\vert2)- $dimensional supersphere$ S^{(2\vert2)}$which is a DeWitt supermanifold over the usual sphere $ S^{2} $ as body. The total manifold is $ S^{(3\vert2)} $with $ (1\vert2)- $dimensional supergroup $ UOSp(2\vert1) $as its supersymmetry group. The structure supergroup is $ U(1) $.
Let $B_{C_{L}} =G^{\infty}(S^{(3\vert2)},C_{L})$ and $A_{C_{L}}= G^{\infty}(S^{(2\vert2)},C_{L})$ denote the graded algebras of $C_{L}$-valued smooth functions on the total super manifold $S^{(3\vert2)}$ and base supermanifold $ S^{(2\vert2)}$ under point-wise multiplication, respectively. Here, $ C_{L} $ is complex Grassmann algebra with $ L $ generators. Similar to the group $ U(1) $, the irreducible representations of the supergroup $ U(1) $ are labeled by an integer $ n $. The elements of $B_{C_{L}} $ can be classified into the right modules,
\begin{equation}
C_{(\pm n)}^{\infty}(S^{(3\vert2)},C_{L}) =\{\varphi_{(\pm n)} :S^{(3\vert2)} \rightarrow C_{L},\quad \varphi_{(\pm n)} (p\cdot\omega)= \omega^{(\pm n)} \cdot \varphi(p)\:, \quad \forall p \in S^{(3\vert2)}\: ,\:\forall\omega\in U(1)\}
\tag{3-2},
\end{equation}
over the pull back of the $A_{C_{L}}$. The left actions of the supergroup $ U(1) $ on $ C_{L} $ are labeled by an integer $ n $ which characterizes the bundle. The Serre-Swan theorem [26] states that for a compact smooth manifold $ M $, there is a complete  equivalence between the category of vector bundles over that manifold and bundle maps, and the category of finitely generated projective modules over the algebra $ C(M) $ of functions over $ M $ and module morphisms. In algebraic $ K $-theory, it is well known that corresponds to these bundles, there are super projectors $ P_{n} $ [25] such that, for the associated super vector bundle 
\begin{equation}
E^{(n)} = S^{(3\vert2)} \times_{U(1)} C_{L} \xrightarrow{\pi} S^{(2\vert2)},
\tag{3-3}
\end{equation}
right $ A_{C_{L}} $-module of sections $\Gamma^{\infty}(S^{(2\vert2)},E^{(n)})$ which is isomorphic with  $G_{(n)}^{\infty}(S^{(3\vert2)}, C_{L})$ is equivalent to the image in the free module $ (A_{C_{L}})^{(2n+1)}=G^{\infty}(S^{(2\vert2)},C_{L}) \otimes C_{L}^{2n+1} $ of a super projector $ P_{n} $, $\Gamma^{\infty}(S^{(2\vert2)},E^{(n)})=P_{n}(A_{C_{L}})^{2n+1} $. The super projector $ P_{n} $ is a super Hermitian operator of super rank $1$ over $C_{L} $
\begin{equation}
P_{n}\in M_{2n+1} (A_{C_{L}}),\quad P_{n}^{2}=P_{n},\quad P_{n}^{\ddagger} =  P_{n},\quad Str P_{n}=1.
\tag{3-4}
\end{equation} 
where $ Str $ is super trace and$ 1 $ is the constant super function.
For the right $ A_{\mathbb{C}} $-module of sections  $ \Gamma^{\infty}(S^{(2\vert2)},E^{(n)}) $ there exist $ n+1 $ super projectors $ P_{1},P_{2},....,P_{n+1} $ having the same super rank $ 1 $. Therefore the free module $ (A_{C_{L}})^{2n+1} $ can be written as a direct sum of the projective $ A_{C_{L}}- $modules,
\begin{equation}
(A_{C_{L}})^{2n+1}=\bigoplus_{\substack{i=1}}^{\substack{2n+1}} P_{i}(A_{C_{L}})^{2n+1}.
\tag{3-5}
\end{equation}
The quantum super Hopf bundle is a $ U(1)- $bundle over the quantum super sphere $ S_{q\mu}^{(2\vert2)} $with standard Podles sphere $ S_{q}^{2} $ as its body and whose total manifold is the super quantum manifold  $ S_{q\mu}^{(3\vert2)} $. The sphere $ S_{q}^{2} $ is a quantum Homogeneous space $ SU_{q}(2) $. Let us denote by $ A(S_{q\mu}^{(3\vert2)}),\,A( S_{q\mu}^{(2\vert2)}) $ and $ A(U(1)) $ the coordinate algebras of the total space $ S_{q\mu}^{(3\vert2)} $, base space $ S_{q\mu}^{(2\vert2)} $ and the fibre $ U(1) $, respectively.
The algebra inclusion $ A( S_{q\mu}^{(2\vert2)})\hookrightarrow  A(S_{q\mu}^{(3\vert2)})$ which is a Hopf-Galois extension is a super quantum Hopf bundle. It is the algebraic version of the first  super quantum Hopf fibration $ S_{q\mu}^{(3\vert2)}\;\xrightarrow{U_{q}(1)}\; S_{q\mu}^{(2\vert2)} $. The quantum associated bundle is given by [28]
\begin{equation}
E_{q}^{(n)}:= \lbrace x \in A(S_{q\mu}^{(3\vert2)}):\:k\cdot x=q^{\frac{n}{2}}x,\:k \in A(U_{q}(1))\rbrace.
\tag{3-6}
\end{equation}
Each $ E_{q}^{(n)} $ is clearly a $ A( S_{q\mu}^{(2\vert2)})- $bi-module and is equivalent to the image in the free module $( A( S_{q\mu}^{(2\vert2)}))^{2n+1} $ of a self-adjoint quantum projector $ P_{q}^{(n)} $ in $ Mat_{2n+1}(A(S_{q\mu}^{(2\vert2)}))  $ (for $ n\geqslant 0 $).
We identify $ E_{q}^{(n)} $ with the left $ A( S_{q\mu}^{(2\vert2)})- $module of sections $( A( S_{q\mu}^{(2\vert2)}))^{2n+1}P_{q}^{(n)} $.

\section{Spin $\frac{1}{2}$ q-deformed superprojectors of the superprojective $ A(S_{q\mu}^{(2\vert2)}) -$module}

According to the Serre-Swan's theorem, in noncommutative geometry, the study of the super principal fibration $S_{q\mu}^{(3\vert2)}\xrightarrow{U(1)} S_{q\mu}^{(2\vert2)}$, replaces with the study of noncommutative finitely generated projective $ A(S_{q\mu}^{(2\vert2)})  -$module of its sections. To build the left and right q-deformed superprojective modules we should construct the q-deformed fuzzy superprojectors of these modules.
The q-deformed superprojectors for left projective module can be written as:
\begin{equation}
P_{[l\pm\frac{1}{2}]_{q}}^{L} = \frac{1}{[2]_{q}}\left\lbrace 1\pm\frac{(\mathbf{\Sigma}\cdot \mathbf{X}+\frac{\mu_{q}}{[2]_{q}})}{\sqrt{\frac{\mu_{q}^{2}}{[2]_{q}^{2}}+1}}\right\rbrace. \tag{4-1}
\end{equation}
where we define inner product as
\begin{equation}
\mathbf{\Sigma}\cdot \mathbf{X}= \sum_{m=0,\pm }(q)^{m}\Sigma_{-m}X_{m}+ \varepsilon_{\alpha\beta}\Sigma_{\alpha}\Theta_{\beta}.
\tag{4-2}
\end{equation}
substituting (2-26) in (4-1) we can write:
\begin{equation}
P_{[l\pm 1/2]_{q}}^{L}= \dfrac{1}{[2]_{q}•}[1\pm \dfrac{[2]_{q}\mathbf{\Sigma}^{q}\cdot \mathbf{L}^{L}+1}{\sqrt{[2l]_{q}[2l+1]_{q}+1}•}]
\tag{4-3}
\end{equation}
which couples left super angular momentum and superspin $ \dfrac{1}{2} $ to its maximum and minimum values $ \l\pm \dfrac{1}{2} $, respectively.
It is easy to see that
\begin{equation}
P_{[l+ \frac{1}{2}]_{q}}^{L} + P_{[l-\frac{1}{2}]_{q}}^{L} = \dfrac{2}{•[2]_{q}}. 
\tag{4-4}
\end{equation}
These are the superprojectors of our left projective $ A(S_{F}^{(2\vert2)})  -$module 
\begin{equation*}
(A(S_{q\mu}^{(2\vert2)}))^{2}=(A(S_{q\mu}^{(2\vert2)}))^{2}P_{[l+ \frac{1}{2}]_{q}}^{L}\oplus (A(S_{q\mu}^{(2\vert2)})^{2}P_{[l- \frac{1}{2}]_{q}}^{L}.
\end{equation*}
Using (4-3) we can define the corresponding super idempotents as:
\begin{equation}
\Gamma_{[l\pm \frac{1}{2}]_{q}}^{L}=[2]_{q}P_{(l\pm \frac{1}{2})}^{L}-1= \pm \dfrac{[2]_{q}\mathbf{\Sigma}^{q}\cdot \mathbf{L}^{L}+1}{\sqrt{[2l]_{q}[2l+1]_{q}+1}•}
\tag{4-5}
\end{equation}
 The q-deformed superprojectors $P_{[l\pm \frac{1}{2}]_{q}}^{R} $ coupling the right super momentum and super spin $ \dfrac{1}{2} $ to its maximum and minimum values $ l\pm \dfrac{1}{2} $, respectively are obtained by changing $ (L_{i}^{L}, L_{\alpha}^{L}) $ to $ (-L_{i}^{R}, -L_{\alpha}^{R}) $ in the above expression
 \begin{equation}
P_{[l\pm 1/2]_{q}}^{R}= \dfrac{1}{[2]_{q}•}[1\mp \dfrac{[2]_{q}\mathbf{\Sigma}^{q}\cdot \mathbf{L}^{R}-1}{\sqrt{[2l]_{q}[2l+1]_{q}+1}•}]
\tag{4-6}
\end{equation}
 These are the q-deformed superprojectors of our right projective $ A(S_{q\mu}^{(2\vert2)})  -$module 
\begin{equation*}
(A(S_{q\mu}^{(2\vert2)}))^{2}=(A(S_{q\mu}^{(2\vert2)}))^{2}P_{[l+ \frac{1}{2}]_{q}}^{R}\oplus (A(S_{q\mu}^{(2\vert2)})^{2}P_{[l- \frac{1}{2}]_{q}}^{R}.
\end{equation*}

 The corresponding q-deformed superidempotents are
 \begin{equation}
\Gamma_{[l\pm \frac{1}{2}]_{q}}^{R}=[2]_{q}P_{(l\pm \frac{1}{2})}^{R}-1= \mp \dfrac{[2]_{q}\mathbf{\Sigma}^{q}\cdot \mathbf{L}^{L}-1}{\sqrt{[2l]_{q}[2l+1]_{q}+1}•}
\tag{4-7}
\end{equation}

\section{Super fuzzy q-deformed Ginsparg-Wilson algebra and its superspin $\frac{1}{2}$ fuzzy q-deformed super Dirac and chirality operators}
 The q-deformed super fuzzy Ginsparg-Wilson algebra $ \mathcal{A}_{q\mu}  $ is the $ \ddagger $ -algebra  over $ C_{L} $, generated by two $ \ddagger $ -invariant super involution $ \Gamma_{q\mu} $ and ${\Gamma^{\prime}_{q\mu}}$:
\begin{equation}
\mathcal{A}_{q\mu} = \langle\Gamma^{q\mu},{\Gamma^{\prime}}^{q\mu}\colon\quad(\Gamma^{q\mu})^{2} =({\Gamma^{\prime}}^{q\mu})^{2}=\textit{I},\quad (\Gamma^{q\mu})^{\ddagger}=\Gamma^{q\mu},\quad ({\Gamma^{\prime}}^{q\mu})^{\ddagger}={\Gamma^{\prime}}^{q\mu}\rangle , 
\tag{5-1}
\end{equation}
each representation of (5-1) is a realization of the q-deformed super Ginsparg-Wilson algebra.
Now, consider the following two elements constructed out of the generators $ \Gamma_{q\mu} $ and ${\Gamma^{\prime}_{q\mu}}  $ of the super fuzzy Ginsparg-Wilson algebra $ \mathcal{A}_{q\mu} $:
\begin{equation*}
\Gamma_{q\mu}^{1}= \frac{1}{[2]_{q}} (\Gamma_{q\mu} + {\Gamma^{\prime}_{q\mu}}) \; , \qquad \qquad {(\Gamma_{q\mu}^{1}})^{\ddagger} = \Gamma_{q\mu}^{1} ,\\
\end{equation*}
\begin{equation}
\Gamma_{q\mu}^{2}= \frac{1}{[2]_{q}} (\Gamma_{q\mu} - {\Gamma^{\prime}_{q\mu}}) \; , \qquad \qquad {(\Gamma_{q\mu}^{2}})^{\ddagger} = \Gamma_{q\mu}^{2} .
\tag{5-2}
\end{equation}
So that, $\Gamma_{q\mu}^{1}$ and $\Gamma_{q\mu}^{2}$ anticommute with each other:
\begin{equation}
\left\lbrace \Gamma_{q\mu}^{1}, \Gamma_{q\mu}^{2}\right\rbrace  = 0.
\tag{5-3}
\end{equation}
Identifying $ \Gamma_{[l\pm \frac{1}{2}]_{q}} ^{L}$ and $ \Gamma_{[l\pm \frac{1}{2}]_{q}} ^{R}$ with $ \Gamma_{q\mu} $ and ${\Gamma^{\prime}_{q\mu}} $, we get:
\begin{equation}
\Gamma_{1}^{\pm}= \pm \frac{\mathbf{\Sigma}^{q} \cdot \mathbf{\mathcal{L}}^{q}+1}{\sqrt{[2l]_{q}[2l+1]_{q}+1}}, \qquad 
\Gamma_{2}^{\pm}= \pm \frac{\mathbf{\Sigma}^{q} \cdot (\mathbf{L}^{L}+\mathbf{L}^{R})}{\sqrt{[2l]_{q}[2l+1]_{q}+1}}.
\tag{5-4}
\end{equation}
Now, let us define the q-deformed super fuzzy Dirac operator on quantum  super fuzzy sphere $ S_{q\mu}^{(2\vert2)} $ as:
\begin{equation}
D_{q\mu}^{\pm }= \sqrt{[2l]_{q}[2l+1]_{q}+1} \Gamma_{1q\mu}^{\pm }= \pm (\mathbf{\Sigma}^{q} \cdot \mathcal{L}^{q}+1)= \pm (\sum_{m=0,\pm}(q)^{m}\Sigma_{-m}^{q} \mathcal{L}_{m}^{q}+ \varepsilon_{\alpha\beta}\Sigma_{\alpha}^{q}\mathcal{L}_{\beta}^{q}+1).
\tag{5-5}
\end{equation}
In the limit case (5-5)becomes the super Dirac operator on the commutative super sphere $ S^{(2\vert2)} $:
\begin{equation}
 \lim_{l \to \infty, q \to 1 } D_{q\mu}^{\pm }= \pm \mathbf{\Sigma} \cdot \mathcal{L} \pm 1= \pm (\Sigma_{i} \mathcal{L}_{i}+ \varepsilon_{\alpha\beta}\Sigma_{\alpha}\mathcal{L}_{\beta}+1).
 \tag{5-6}
\end{equation}
where $ \mathcal{L}_{i} $and $ \mathcal{L}_{\alpha} $ are super angular momentum operators on super sphere $ S^{(2\vert2)} $
\begin{equation}
\mathcal{L}_{i}=-i\varepsilon_{ijk}x_{j}\dfrac{\partial}{•\partial x_{k}}+\dfrac{1}{2•}\theta_{\alpha}(\sigma_{i})_{\alpha\beta}\dfrac{\partial}{\partial \theta_{\beta}•}\tag{5-7}
\end{equation}
and
\begin{equation}
\mathcal{L}_{\alpha}= \dfrac{1}{2•}x_{i}(\varepsilon \sigma_{i})_{\alpha\beta}\dfrac{\partial}{\partial \theta_{\beta}•}- \dfrac{1}{2•}\theta_{\beta}(\sigma_{i})_{\beta\alpha}\dfrac{\partial}{\partial x_{i}•}\tag{5-8}
\end{equation}
Also, we can define q-deformed super chirality operator on $ S_{q\mu}^{(2\vert2)} $ as $ \gamma_{q\mu}^{\pm }= \Gamma_{2q\mu}^{\pm } $ which in the commutative limit it becomes:
\begin{equation}
 \lim_{l\to \infty, q \to 1} \gamma_{q\mu}^{\pm }=\pm \Sigma\cdot x=\pm( \Sigma_{i}x_{i}+ \varepsilon_{\alpha\beta}\Sigma_{\alpha}\theta_{\beta}).
 \tag{5-9}
\end{equation}
Also, it is easy to see that 
\begin{equation}
 \lim_{l\to \infty, q\to 1} \lbrace D_{q\mu}^{\pm },\gamma_{q\mu}^{\pm }\rbrace=0
 \tag{5-10}
\end{equation}
which we expect from Dirac and chirality operators on $ S^{(2\vert2)} $.

\section{Super fuzzy gauged q-deformed Dirac operator ( no instanton fields)}
Let us denote by $ A^{L} $ the super connection $ 1- $form associated with the q-deformed superprojector $ P $
\begin{equation}
 A^{L}\in End_{B_{C_{L}}}( G^{\infty}(UOSp_{q}(2\vert1)),C_{L})\otimes_{B_{C_{L}}}\Omega^{1}( UOSp_{q}(2\vert1)),C_{L})
 \tag{6-1}
\end{equation}
The components of this super $ U(1) $gauged field according to our q-deformed super Hopf fiberation are given by 
\begin{equation}
 A= \sum_{m=0,\pm}(q)^{m}dx_{-m}A_{m}+ d\theta_{\alpha}A_{\alpha}
 \tag{6-2}
\end{equation}
The $ \ddagger- $invariant super fuzzy gauge field $ A^{L} $ acts on $ \xi = ( \xi_{1} ,.....,\xi_{k}),\xi_{i} \in  S^{(2\vert2)}_{q\mu}  (4l + 1) $ as:
\begin{equation}
(A_{i,\alpha}^{L}\xi)_{m}=(A_{i,\alpha})_{mn}\xi_{n}.
\tag{6-3}
\end{equation}
The $ \ddagger $-invariant condition on $ A_{i,\alpha}^{L} $ is:
\begin{equation}
(A_{i}^{L})^{\ddagger} = A_{i}^{L}, \quad (A_{\alpha}^{L})^{\ddagger}= -(\sigma_{1})_{\alpha\beta}A_{\beta}.
\tag{6-4}
\end{equation}
The super fuzzy gauge field $ A^{L} $ on the commutative super sphere $ S^{(2\vert2)} $ becomes a commutative field $ \mathbf{a}$ and its components $ a_{i,\alpha} $satisfies the following condition: 
\begin{equation}
\mathbf{x}\cdot \mathbf{a}=x_{i}a_{i}+ \varepsilon_{\alpha\beta}\theta_{\alpha}a_{\beta}= 0.
\tag{6-5}
\end{equation}
We need a condition to get the above result for large $ l $. One of the conditions of such a nature is:
\begin{equation}
(\mathbf{L}^{L} + \mathbf{A}^{L}) \cdot (\mathbf{L}^{L} + \mathbf{A}^{L})= \mathbf{L}^{L} \cdot \mathbf{L}^{L} = \sum_{m=0,\pm}(q)^{m}L_{-m}L_{m}+ \varepsilon_{\alpha\beta}L_{\alpha}L_{\beta}= \dfrac{•[2l]_{q}[2l+1]_{q} }{[2]_{q}^{2}•}.
\tag{6-6}
\end{equation}
The expansion of (6-6) is:
\begin{equation}
\sum_{m=0,\pm}(q)^{m}(A_{-m}^{L}L_{m}^{L}+ L_{-m}^{L}A_{m})+ \varepsilon_{\alpha\beta}(L_{\alpha}^{L}A_{\beta}^{L}+ A_{\alpha}^{L} L_{\beta}^{L})+A^{L}\cdot A^{L}=0
\tag{6-7}
\end{equation}
When the parameter $ l $ tends to infinity and $ q\rightarrow 1 $, $ \dfrac{A_{i,\alpha}^{L}}{l} $ tends to zero.
 Also, in this limit $ L_{i}^{L} $, $ L_{\alpha}^{L} $ and $ A_{i,\alpha}^{L} $ tends to $ x_{i}$, $ \theta_{\alpha} $and $ a $, respectively . So we have the condition $ \mathbf{x} \cdot \mathbf{a}=0 $ on the commutative supersphere.\\
Now, we can introduced the q-deformed super gauged Ginsparg-Wilson system as follow: We can set:
\begin{equation}
\Gamma^{\pm }_{q\mu}(\mathbf{A}^{L}) = \pm \dfrac{[2]_{q}\mathbf{\Sigma}^{q}\cdot (\mathbf{L}^{L} + \mathbf{A}^{L})+1}{|[2]_{q}\mathbf{\Sigma}^{q}\cdot (\mathbf{L}^{L} + \mathbf{A}^{L})+1|},\quad \Gamma^{' \pm }_{q\mu}(\mathbf{A}^{L}) = \Gamma^{' \pm }(0)=\mp \dfrac{[2]_{q}\mathbf{\Sigma}^{q}\cdot \mathbf{L}^{R}-1}{|[2]_{q}\mathbf{\Sigma}^{q}\cdot \mathbf{L}^{L}-1|}
\tag{6-8}
\end{equation}
It is an involutory and $ \ddagger $-invariant operator:
\begin{equation}
\Gamma_{q\mu}(\mathbf{A}^{L})^{2} = 1 ,\qquad \Gamma_{q\mu}(\mathbf{A}^{L})^{\ddagger} = \Gamma_{q\mu}(\mathbf{A}^{L}).
\tag{6-9}
\end{equation}
The gauged involution (6-8), reduces to (4-5) for zero $ \mathbf{A}^{L}$. We put $ \Gamma_{q\mu} = \Gamma_{q\mu}(\mathbf{A}^{L}=0) $.\\ Also, we can define the second gauged involution as:
\begin{equation}
\Gamma_{q\mu}^{'}(\mathbf{A}^{L}) = \Gamma_{q\mu}^{'}(0)=\Gamma_{q\mu}^{'}.
\tag{6-10}
\end{equation}
We put $ \Gamma_{q\mu}^{'}=\Gamma_{q\mu}^{'}(A^{L}=0) $. Notice that, the operators $ \mathbf{L}^{L,R} $  do not have continuum limit as their squares $ \dfrac{[2l]_{q}[2l+1]_{q}•}{•[2]_{q}^{2}}$ diverege as $ l $ tends to infinity. In contrast, $\boldsymbol{\mathcal{L}} $ and $ \mathbf{A}^{L} $ do have continuum limits.\\
It is easy to see that up to the first order (6-8) becomes:
\begin{equation}
\Gamma_{q\mu}^{\pm} (\mathbf{A}^{L}) = \pm \dfrac{[2]_{q} \mathbf{\Sigma}^{q}\cdot (\mathbf{L}^{L} + \mathbf{A}^{L})+1}{\sqrt{[2l]_{q}[2l+1]_{q}+1}}.
\tag{6-11}
\end{equation}
and
\begin{equation}
\Gamma_{q\mu}^{' \pm } = \mp\dfrac{[2]_{q} \mathbf{\Sigma}^{q}\cdot \mathbf{L}^{R} -1 }{\sqrt{[2l]_{q}[2l+1]_{q}+1}}.
\tag{6-12}
\end{equation}
Using (6-11)and (6-12) we can construct the following $ \ddagger- $invariant operators:
\begin{equation}
\begin{split}
\Gamma_{1q\mu}^{\pm }(A^{L}) = \frac{1}{[2]_{q}} (\Gamma_{q\mu}^{\pm }(A^{L}) + \Gamma_{q\mu}^{\prime \pm}) \; , \qquad \qquad (\Gamma_{1q\mu}^{\pm })^{\ddagger} = \Gamma_{1q\mu}^{\pm } ,\\
\Gamma_{2q\mu}^{\pm }(A^{L}) = \frac{1}{[2]_{q}} (\Gamma_{q\mu}^{\pm }(A^{L}) - \Gamma_{q\mu}^{\prime \pm }) \; , \qquad \qquad (\Gamma_{2q\mu}^{\pm })^{\ddagger} =  \Gamma_{2q\mu}^{\pm } .
\end{split}
\tag{6-13}
\end{equation}
Now, let us define the q-deformed gauged super fuzzy Dirac and chirality operators on quantum super fuzzy sphere $ S_{q\mu}^{(2\vert2)} $as:
\begin{equation}
D_{q\mu}^{\pm }(\mathbf{A}^{L})=\sqrt{[2l]_{q}[2l+1]_{q}+1}\Gamma_{1q\mu}^{\pm }(\mathbf{A}^{L})=\pm (\mathbf{\Sigma}^{q} \cdot (\mathcal{L}^{q}+\mathbf{A}^{L})+ 1),\qquad 
\tag{6-14}
\end{equation}
and for the q-deformed super chirality operator:
\begin{equation}
\gamma_{q\mu}^{\pm }(\mathbf{A}^{L})=\Gamma_{2q\mu}^{\pm }(\mathbf{A}^{L})= \pm \dfrac{[2]_{q} \mathbf{\Sigma}^{q}\cdot (\mathbf{L}^{L}+\mathbf{L}^{R} + \mathbf{A}^{L})}{\sqrt{[2l]_{q}[2l+1]_{q}+1}}.
\tag{6-15}
\end{equation}
In the third commutative limit, (6-14) and (6-15) become:
\begin{equation}
 \lim_{l \to \infty, q \to 1} D_{q\mu}^{\pm }(A^{L})=\pm (\mathbf{\Sigma} \cdot (\mathcal{L}+A^{^{L}})+1),\qquad  \lim_{l \to \infty, q\to 1}\gamma_{q\mu}^{\pm }(A^{L}) =\pm \mathbf{\Sigma} \cdot \mathcal{\mathbf{x}}.
 \tag{6-16}
\end{equation}
These are the correct gauged super Dirac and chirality operators on commutative super sphere$ S^{(2\vert2)} $.

\section{Instanton coupling}
As we mentioned in section $3$, according to the Serre-Swan's theorem, in noncommutative geometry, the study of the quantum super principal fibration $S_{q\mu}^{(3\vert2)}\xrightarrow{U(1)} S_{q\mu}^{(2\vert2)}$, replaces with the study of noncommutative finitely generated projective $ A(S_{q\mu}^{(2\vert2)})  -$module of its sections. To build the projective module, let introduce  $\mathbb{C}^{4t+1}$ carrying the $ t $-representation of quantum super angular momentum of $ U_{q}(uosp(2\vert1))$.
Here, the algebra $ U_{q}(uosp(2\vert1)) $ is generated by elements $ T_{0} $ and $ T_{\alpha} $ satisfying the following relation:
\begin{equation}
[T_{0}, T_{\pm }]= \pm T_{\pm }, \quad  \lbrace T_{+}, T_{-} \rbrace = [2T_{0}]_{q^{1/2}}.
\tag{7-1}
\end{equation}
They also satisfy the relations (2-33)-(2-37).
 Let $ P_{[(l+ t)\pm 1/2]_{q}}^{L} $ be the  super projector coupling left super angular momentum operator $ \mathbf{L}^{L}$ with $ \mathbf{T} $ to produce maximum angular momentum $ l+t $. We know that the image of a  projector on a free module is a projective module. Then, as $ Mat(4l+1)^{4t+1}=Mat (4l+1)\otimes \mathbb{C}^{4t+1} $ is a free module, therefore, $  P_{[(l\pm t)\pm 1/2]_{q}}^{L} Mat(4l+1)^{4t+1} $ is the fuzzy version of $ U(1) $ bundle on $S_{q\mu}^{(2\vert2)}$. Also, we can use the quantum super projector $ P_{[(l- t)\pm 1/2]_{q}}^{L} $ to produce the projective module $ P_{[(l\pm t)\pm 1/2]_{q}}^{L} Mat (4l+1)^{4t+1}$ to introduce the least angular momentum $ (l-t) $.\\ The superspin $ 1/2 $quantum super fuzzy projectors $P_{[(l\pm t)\pm 1/2]_{q}}^{L}$ corresponding to $ (l\pm t)$-representations of $ U_{q}(uosp(1\vert2)) $ can be written as:
\begin{equation}
P_{[(l\pm t)\pm 1/2]_{q}}^{L}= \dfrac{1}{[2]_{q}•}[1\pm \dfrac{[2]_{q}\mathbf{\Sigma}^{q}\cdot (\mathbf{L}^{L}+ \mathbf{T})+1}{\sqrt{[2(l\pm t)]_{q}[2(l\pm t)+1]_{q}+1}•}],\quad P_{[(l\pm t)\pm 1/2]_{q}}^{L\ddagger}= P_{[(l\pm t)\pm 1/2]_{q}}^{L},
\tag{7-2}
\end{equation}
\begin{equation}
Mat(4l+1)\otimes\mathbb{C}^{4t+1}=(Mat(4l+1)\otimes\mathbb{C}^{4t+1})P_{[(l+ t)\pm 1/2]_{q}}^{L}\oplus  (Mat(4l+1)\otimes \mathbb{C}^{4t+1})P_{[(l- t)\pm 1/2]_{q}}^{L}.
\tag{7-3}
\end{equation}
To set the quantum super fuzzy Ginsparg-Wilson system in super instanton sector, we choose the following $ \ddagger-$invariant involution $ \Gamma_{q\mu} $ for the highest and lowest super weights $ l\pm t $:
\begin{equation}
\Gamma_{q\mu}^{\pm L}(\mathbf{T})=[2]_{q} P_{[(l\pm t)\pm 1/2]_{q}}^{L}-1=\pm \dfrac{[2]_{q}\mathbf{\Sigma}^{q}\cdot (\mathbf{L}^{L}+ \mathbf{T})+1}{\sqrt{[2(l\pm t)]_{q}[2(l\pm t)+1]_{q}+1}•},
\tag{7-4}
\end{equation}
We choose $ \Gamma_{q\mu}^{'} $ as in (6-12). It is clear that $ \Gamma_{q\mu}^{\pm}(T=0)=\Gamma_{q\mu}^{\pm} $. On the module $( Mat(4l+1)^{4t+1}\otimes\mathbb{C}^{2})P_{[(l\pm t)\pm 1/2]_{q}}^{L} $ we have:
\begin{equation}
(\mathbf{L}^{L}+\mathbf{T})^{2} = \dfrac{[2(l\pm t)]_{q}[2(l\pm t)+1]_{q}}{•[2]_{q}^{2}}.
\tag{7-5}
\end{equation}
The quantum superprojectors $P_{[(l\pm t)\pm \frac{1}{2}]_{q}}^{R} $ coupling the right super momentum, super instanton and super spin $ \dfrac{1}{2} $ to its maximum and minimum values $ (l\pm t)\pm \dfrac{1}{2} $, respectively are obtained by changing $ (L_{i}^{L}, L_{\alpha}^{L}) $ to $ (-L_{i}^{R}, -L_{\alpha}^{R}) $ in the above expression
 \begin{equation}
P_{[(l\pm t)\pm 1/2]_{q}}^{R}= \dfrac{1}{[2]_{q}•}[1\pm \dfrac{[2]_{q}\mathbf{\Sigma}^{q}\cdot (-\mathbf{L}^{R}+ \mathbf{T})+1}{\sqrt{[2(l\pm t)]_{q}[2(l\pm t)+1]_{q}+1}•}]
\tag{7-6}
\end{equation}
 These are the quantum superprojectors of our right projective $ A(S_{q\mu}^{(2\vert2)})  -$module 
\begin{equation*}
(A(S_{q\mu}^{(2\vert2)}))^{2}=(A(S_{q\mu}^{(2\vert2)}))^{2}P_{[(l\pm t)+ \frac{1}{2}]_{q}}^{R}\oplus (A(S_{q\mu}^{(2\vert2)}))^{2}P_{[(l\pm t)- \frac{1}{2}]_{q}}^{R}.
\end{equation*}
 The corresponding quantum superidempotents are
 \begin{equation}
\Gamma_{q\mu}^{\pm R}(\mathbf{T})=[2]_{q} P_{[(l\pm t)\pm 1/2]_{q}}^{R}-1=\pm \dfrac{[2]_{q}\mathbf{\Sigma}^{q}\cdot (-\mathbf{L}^{R}+ \mathbf{T})+1}{\sqrt{[2(l\pm t)]_{q}[2(l\pm t)+1]_{q}+1}•}
\tag{7-7}
\end{equation}
Now, we can introduce our quantum super fuzzy Ginsparg-Wilson system in super instanton sector as:
\begin{equation}
\mathcal{A}_{q\mu}^{\pm}(\mathbf{T})=\langle\; \Gamma_{q\mu}^{\pm}(\mathbf{T}), \Gamma_{q\mu}^{'} : \Gamma_{q\mu}^{\pm^{^{2}}}(\mathbf{T})=\Gamma_{q\mu}^{'^{2}}=1,\quad\Gamma_{q\mu}^{\pm \ddagger}(\mathbf{T})=\Gamma_{q\mu}^{\pm}(\mathbf{T}),\quad\Gamma_{q\mu}^{'\ddagger}= \Gamma_{q\mu}^{'} \rangle.
\tag{7-8}
\end{equation}
Now, consider the following two elements constructed out of the generators $ \Gamma_{q\mu}(\mathbf{T}) $ and ${\Gamma_{q\mu}^{\prime}}  $ of the quantum super fuzzy Ginsparg-Wilson algebra $ \mathcal{A}_{q\mu}(\mathbf{T}) $:
\begin{equation}
\begin{split}
\Gamma_{1q\mu}^{\pm}(\mathbf{T})= \frac{1}{[2]_{q}} (\Gamma_{q\mu}^{\pm}(\mathbf{T}) + {\Gamma_{q\mu}^{\prime}}) \; , \qquad \qquad {(\Gamma_{1q\mu}})^{\ddagger} = \Gamma_{1q\mu} ,\\
\Gamma_{2q\mu}^{\pm}(\mathbf{T}) = \frac{1}{[2]_{q}} (\Gamma_{q\mu}^{\pm}(\mathbf{T}) - {\Gamma_{q\mu}^{\prime}}) \; , \qquad \qquad {(\Gamma_{2q\mu}})^{\ddagger} = \Gamma_{2q\mu}.
\end{split}
\tag{7-9}
\end{equation}
Identifying $ \Gamma_{[(l\pm t)\pm \frac{1}{2}]_{q}} ^{L}$ and $ \Gamma_{[(l\pm t)\pm \frac{1}{2}]_{q}} ^{R}(T=0)$ with $ \Gamma $ and ${\Gamma^{\prime}} $, it is easy to compute $ \Gamma_{q\mu}^{\pm 1} $ and $ \Gamma_{q\mu}^{\pm 2} $. Now we can define q-deformed super pseudo Dirac  and chirality operators as
\begin{equation*}
D_{q\mu}^{\pm}(\mathbf{T})= \sqrt{[2(l\pm t)]_{q}[2(l\pm t)+1]_{q}+1}•\sqrt{[2l]_{q}[2l+1]_{q}+1}•\Gamma_{1q\mu}^{\pm}(\mathbf{T})
\end{equation*}
\begin{equation}
\gamma_{q\mu}^{\pm}(\mathbf{T})= \sqrt{[2(l\pm t)]_{q}[2(l\pm t)+1]_{q}+1}•\sqrt{[2l]_{q}[2l+1]_{q}+1}•\Gamma_{2q\mu}^{\pm}(\mathbf{T})
\tag{7-10}
\end{equation}
 which in the third commutative limit become:
\begin{equation}
 \lim_{l \to \infty, q \to 1} D_{q\mu}^{\pm }(\mathbf{T})=\pm (\mathbf{\Sigma} \cdot (\mathcal{L}+T)+ 1),\qquad  \lim_{l \to \infty, q\to 1}\gamma_{F}^{\pm }(\mathbf{T}) =\pm \mathbf{\Sigma} \cdot \mathcal{\mathbf{x}}.
 \tag{7-11}
\end{equation}
These are the correct q-deformed super Dirac and chirality operators on commutative supersphere $ S^{(2\vert2)} $ in the instanton sector.
It is obvious that the Dirac operator (7-10) is $ \ddagger $-invariant:
\begin{equation}
D_{q\mu}^{(\pm )^{\ddagger}}(T)= D_{q\mu}^{(\pm )}(T).
\tag{7-12}
\end{equation}
which we expect from commutative Dirac operator in instanton sector.

\section{Gauging the super fuzzy Dirac operator in instanton sector}
The derivation $\mathcal{L}_{i,\alpha} $ dose not commute with the projectors $ P_{q\mu}^{(l\pm t)} $ and then has no action on the modules $Mat(4l+1)P_{q\mu}^{(l \pm t)}$. But $J_{i,\alpha}=\mathcal{L}_{i,\alpha}+T_{i,\alpha}$ does commute with $ P_{q\mu}^{(l\pm t)}$. Here, $ J_{i,\alpha} $ has been considered as the total quantum super angular momentum. Now, we need to gauge $J_{i,\alpha} $. When $ T=0$, the super gauge fields $ A_{i,\alpha} $ were function of $ L_{i,\alpha}^{L}$. Here, we consider $ A_{i,\alpha}^{L} $ to be a functions of $ \mathbf{L}^{L}+\mathbf{T} $, because $ A_{i,\alpha}^{L} $ dose not commute with $ P_{q\mu}^{(l\pm t)} $. Let us introduce the super covariant derivative as:
\begin{equation}
\nabla_{i}=J_{i}+A_{i}^{L} ,\quad \nabla_{\alpha}=J_{\alpha}+A_{\alpha}^{L}.
\tag{8-1}
\end{equation}
In this case the limiting transversality of $ \mathbf{L}^{L}+\mathbf{T} $ can be guaranteed by imposing the condition:
\begin{equation}
(\mathbf{L}^{L}+\mathbf{A}^{L}+\mathbf{T})\cdot(\mathbf{L}^{L}+\mathbf{A}^{L}+\mathbf{T})=(\mathbf{L}^{L}+\mathbf{T})\cdot(\mathbf{L}^{L}+\mathbf{T})= \dfrac{[2(l\pm t)]_{q}[2(l\pm t)+1]_{q}}{•[2]_{q}^{2}},
\tag{8-2}
\end{equation}
The expansion of (8-2) is:
\begin{equation}
(\mathbf{L}^{L}+\mathbf{T})\cdot \mathbf{A}^{L} +\mathbf{A}^{L}\cdot(\mathbf{L}^{L}+\mathbf{T})+ \mathbf{A}^{L}\cdot \mathbf{A}^{L}=0.
\tag{8-3}
\end{equation}
When the parameter $ l $ tends to infinity, $ \dfrac{A_{i,\alpha}^{L}}{l} $ and $ \dfrac{\mathbf{T}}{l•} $ tend to zero and $ (L_{i}^{L}, L_{\alpha}^{L}) $ tend to $ (x_{i}, \theta_{\alpha}) $. Then, for large $ l $, the (8-3) tends to the condition $ \mathbf{x}\cdot \mathbf{a}=0 $.
In this case the quantum super left projectors are given by
\begin{equation}
P_{[(l\pm t)\pm 1/2]_{q}}^{L}= \dfrac{1}{[2]_{q}•}[1\pm \dfrac{[2]_{q}\mathbf{\Sigma}^{q}\cdot (\mathbf{L}^{L}+ \mathbf{T}+ \mathbf{A}^{L})+1}{\sqrt{[2(l\pm t)]_{q}[2(l\pm t)+1]_{q}+1}•}],\quad P_{[(l\pm t)\pm 1/2]_{q}}^{L\ddagger}= P_{[(l\pm t)\pm 1/2]_{q}}^{L},
\tag{8-4}
\end{equation}
and the corresponding involutions have the form
 \begin{equation}
\Gamma_{q\mu}^{\pm L}(\mathbf{T,A^{L}})=[2]_{q} P_{[(l\pm t)\pm 1/2]_{q}}^{L}-1=\pm \dfrac{[2]_{q}\mathbf{\Sigma}^{q}\cdot (\mathbf{L}^{L}+ \mathbf{T}+ \mathbf{A}^{L})+1}{\sqrt{[2(l\pm t)]_{q}[2(l\pm t)+1]_{q}+1}•},
\tag{8-5}
\end{equation}
We choose $ \Gamma_{q\mu}^{'} $ as in (6-12).
The quantum superprojectors $P_{[(l\pm t)\pm \frac{1}{2}]_{q}}^{R} $ coupling the right super momentum, super instanton and super spin $ \dfrac{1}{2} $ to its maximum and minimum values $ (l\pm t)\pm \dfrac{1}{2} $, respectively are obtained by changing $ (L_{i}^{L}, L_{\alpha}^{L}) $ to $ (-L_{i}^{R}, -L_{\alpha}^{R}) $ in the above expression
 \begin{equation}
P_{[(l\pm t)\pm 1/2]_{q}}^{R}= \dfrac{1}{[2]_{q}•}[1\pm \dfrac{[2]_{q}\mathbf{\Sigma}^{q}\cdot (-\mathbf{L}^{R}+ \mathbf{T}+ \mathbf{A}^{L})+1}{\sqrt{[2(l\pm t)]_{q}[2(l\pm t)+1]_{q}+1}•}]
\tag{8-6}
\end{equation}
and the corresponding idempotents can be given by
\begin{equation}
\Gamma_{q\mu}^{\pm R}(\mathbf{T, A^{L}})=[2]_{q} P_{[(l\pm t)\pm 1/2]_{q}}^{R}-1=\pm \dfrac{[2]_{q}\mathbf{\Sigma}^{q}\cdot (-\mathbf{L}^{R}+ \mathbf{T}+ \mathbf{A}^{L})+1}{\sqrt{[2(l\pm t)]_{q}[2(l\pm t)+1]_{q}+1}•}
\tag{8-7}
\end{equation}
Now, we can construct the gauged quantum super fuzzy Ginsparg-Wilson system in instanton sector and its corresponding quantum super fuzzy Dirac and chirality operators as follow: 
\begin{equation}
\mathcal{A}^{\pm}( \mathbf{T},\mathbf{A}^{L})= \langle \Gamma^{\pm}(\mathbf{T,A^{L}}), \Gamma^{'} : \Gamma^{\pm^{^{2}}}(\mathbf{T,A^{L}})=\Gamma^{'^{2}}=1,\quad\Gamma^{\pm \ddagger}=\Gamma^{\pm} ,\quad\Gamma^{'\ddagger}= \Gamma^{'} \rangle.
\tag{8-8}
\end{equation}
Now, consider the following two elements constructed out of the generators $ \Gamma_{q\mu}(\mathbf{T,A^{L}}) $ and ${\Gamma_{q\mu}^{\prime}}  $ of the quantum super fuzzy Ginsparg-Wilson algebra $ \mathcal{A}_{q\mu}(\mathbf{T,A^{L}}) $:
\begin{equation}
\begin{split}
\Gamma_{1q\mu}^{\pm}(\mathbf{T,A^{L}})= \frac{1}{[2]_{q}} (\Gamma_{q\mu}^{\pm}(\mathbf{T,A^{L}}) + {\Gamma_{q\mu}^{\prime}}) \; , \qquad \qquad {(\Gamma_{1q\mu}})^{\ddagger} = \Gamma_{1q\mu},\\
\Gamma_{2q\mu}^{\pm}(\mathbf{T,A^{L}}) = \frac{1}{[2]_{q}} (\Gamma_{q\mu}^{\pm}(\mathbf{T,A^{L}}) - {\Gamma_{q\mu}^{\prime}}) \; , \qquad \qquad {(\Gamma_{2q\mu}})^{\ddagger} = \Gamma_{2q\mu}.
\end{split}
\tag{8-9}
\end{equation}
Identifying $ \Gamma_{[(l\pm t)\pm \frac{1}{2}]_{q}} ^{L}$ and $ \Gamma_{[(l\pm t)\pm \frac{1}{2}]_{q}} ^{R}(T=0)$ with $ \Gamma $ and ${\Gamma^{\prime}} $, it is easy to compute $ \Gamma_{q\mu}^{\pm 1} $ and $ \Gamma_{q\mu}^{\pm 2} $. Now we can define q-deformed super pseudo Dirac  and chirality operators as
\begin{equation*}
D_{q\mu}^{\pm}(\mathbf{T,A^{L}})= \sqrt{[2(l\pm t)]_{q}[2(l\pm t)+1]_{q}+1}•\sqrt{[2l]_{q}[2l+1]_{q}+1}•\Gamma_{1q\mu}^{\pm}(\mathbf{T})
\end{equation*}
\begin{equation}
\gamma_{q\mu}^{\pm}(\mathbf{T,A^{L}})= \sqrt{[2(l\pm t)]_{q}[2(l\pm t)+1]_{q}+1}•\sqrt{[2l]_{q}[2l+1]_{q}+1}•\Gamma_{2q\mu}^{\pm}(\mathbf{T})
\tag{8-10}
\end{equation}
 which in the third commutative limit become:
\begin{equation}
 \lim_{l \to \infty, q\to 1} D_{q\mu}^{\pm }(\mathbf{T}, \mathbf{A}^{L})=\pm(\mathbf{\Sigma} \cdot (\mathcal{L}+ \mathbf{T}+ \mathbf{A}^{L})+ 1),\qquad  \lim_{l \to \infty, q\to 1}\gamma_{q\mu}^{\pm }(\mathbf{T}, \mathbf{A}^{L}) =\pm \mathbf{\Sigma} \cdot \mathcal{\mathbf{x}}.
 \tag{8-11}
\end{equation}
which we expect from commutative super gauged Dirac and chirality operators in instanton sector on $ S^{(2\vert2)} $.

\section{Conclusion}
\textbf{}
In this paper, using the quantum super projectors and idempotents of the finitely generated quantum super projective $ A(S_{q\mu}^{(2\vert2)})  -$module it has been constructed the generators of the quantum super gauged fuzzy Ginsparg-Wilson algebra in instanton sector. It has been constructed q-deformed super gauged fuzzy Dirac operator in instanton sector using the quantum super fuzzy Ginsparg-Wilson algebra. The importance of this Dirac operator is that it has correct commutative limit.

\end{document}